\documentclass[iop]{emulateapj}
\usepackage{amsmath}
\usepackage{amssymb}
\usepackage{mathbbol}
\usepackage{dsfont}

\DeclareGraphicsExtensions{.eps,.eps.gz,.epsi}

\shorttitle{Photometric Metallicity with a Well-designed Filter and Gaia XP Spectra}
\shortauthors{Xiao et al.}
\begin{document}

\title{Filter Design for Estimation of Stellar Metallicity: Insights from Experiments with Gaia XP Spectra}

\author
{Kai Xiao\altaffilmark{1},
Bowen Huang\altaffilmark{2, 3},
Yang Huang\altaffilmark{1, 4},
Haibo Yuan\altaffilmark{2, 3},
Timothy C. Beers\altaffilmark{5},
Jifeng Liu\altaffilmark{4, 1, 2}
Maosheng Xiang\altaffilmark{4, 2},
Xue Lu\altaffilmark{2, 3},
Shuai Xu\altaffilmark{2, 3},
Lin Yang\altaffilmark{6},
Chuanjie Zheng\altaffilmark{4, 1},
Zhirui Li\altaffilmark{4, 1},
Bowen Zhang\altaffilmark{4, 1},
Ruifeng Shi\altaffilmark{1},
}
\altaffiltext{1}{School of Astronomy and Space Science, University of Chinese Academy of Sciences, Beijing 100049, People's Republic of China; email: huangyang@ucas.ac.cn}
\altaffiltext{2}{Institute for Frontiers in Astronomy and Astrophysics, Beijing Normal University, Beijing, 102206, China; email: yuanhb@bnu.edu.cn}
\altaffiltext{3}{Department of Astronomy, Beijing Normal University, Beijing, 100875, People's Republic of China}
\altaffiltext{4}{CAS Key Lab of Optical Astronomy, National Astronomical Observatories, Chinese Academy of Sciences, Beijing 100012, People's Republic of China}
\altaffiltext{5}{Department of Physics and Astronomy and JINA Center for the Evolution of the Elements (JINA-CEE), University of Notre Dame, Notre Dame, IN 46556, USA}
\altaffiltext{6}{Department of Cyber Security, Beijing Electronic Science and Technology Institute, Beijing, 100070, China}

\journalinfo{Accepted to ApJL}
\submitted{Received May 09, 2024; Revised May, 29, 2024; Accepted May, 30, 2024}

\begin{abstract}

We search for an optimal filter design for the estimation of stellar metallicity, based on synthetic photometry from Gaia XP spectra convolved with a series of filter-transmission curves defined by different central wavelengths and bandwidths.  Unlike previous designs based solely on maximizing metallicity sensitivity, we find that the optimal solution provides a balance between the sensitivity and uncertainty of the spectra. With this optimal filter design, the best precision of metallicity estimates for relatively bright ($G \sim 11.5$) stars is excellent, $\sigma_{\rm [Fe/H]} = 0.034$\,dex for FGK dwarf stars, superior to that obtained utilizing custom sensitivity-optimized filters (e.g., SkyMapper\,$v$).  By selecting hundreds of high-probabability member stars of the open cluster M67, our analysis reveals that the intrinsic photometric-metallicity scatter of these cluster members is only 0.036\,dex, consistent with this level of precision. Our results clearly demonstrate that the internal precision of photometric-metallicity estimates can be extremely high, even providing the opportunity to perform chemical tagging for very large numbers of field stars in the Milky Way. This experiment shows that it is crucial to take into account uncertainty alongside the sensitivity when designing filters for measuring the stellar metallicity and other parameters.
\end{abstract}

\keywords{Fundamental parameters of stars (555); Metallicity (1031); Astronomy data analysis (1858); Photometry (1234)}

\section{Introduction} \label{sec:intro}

Accurate and precise determinations of stellar parameters for large samples of stars plays a vital role in numerous fields of astronomy, including stellar physics and various Galactic studies.  At present these parameters, especially metallicity, are mainly derived via spectroscopy. In the past decades, stellar parameters for over ten million stars have been obtained from massive ground-based spectroscopic surveys, such as the Sloan Extension for Galactic Understanding and Exploration \citep[SEGUE;][]{2009AJ....137.4377Y,2022ApJS..259...60R}, the Radial Velocity Experiment \citep[RAVE;][]{2006AJ....132.1645S}, the Apache Point Observatory Galactic Evolution Experiment \citep[APOGEE;][]{2017AJ....154...94M}, the Large Sky Area Multi-Object Fiber Spectroscopic Telescope \citep[LAMOST;][]{2012RAA....12.1197C,2012RAA....12..723Z}, the Galactic Archaeology with HERMES \citep[GALAH;][]{2015MNRAS.449.2604D}, and the Dark Energy Spectroscopic Instrument \citep[DESI;][]{2016arXiv161100036D}. Numerous additional massive spectroscopic surveys are soon to begin (e.g., MOONS, WEAVE, 4MOST, PFS; \citealt{2014SPIE.9147E..0NC,2016ASPC..507...97D,2019Msngr.175....3D,
2016SPIE.9908E..1MT}), which will significantly expand these numbers.
However, the total stellar targets observed by such present and future spectroscopic surveys will still lag far behind the numbers  of stars with astrometric parameter estimates measured by the Gaia mission \citep{2023A&A...674A...1G}.

The efficiency of obtaining stellar-parameter estimates can be significantly enhanced through a photometric approach utilizing specially designed filter systems \citep[e.g.,][]{2022ApJ...925..164H}.
The best known is the Str{\"o}mgren $uvby$ System \citep{1963QJRAS...4....8S}, which achieved remarkable success in deriving stellar-atmospheric parameters for the Geneva-Copenhagen survey of the Solar Neighborhood \citep[GCS;][]{2004A&A...418..989N}.  In addition to this system, \citet{2011PASP..123..789B} designed the SkyMapper filter set by inserting a $v$-band filter between the SDSS-like $u$- and $g$-bands to obtain high-sensitivity to metallicity, and a Str{\"o}mgren-like $u$-band to provide sensitivity to both temperature (for hot stars) and surface gravity (for A/F/G/K-type stars). This system has been adopted by the SkyMapper Southern Survey \citep[SkyMapper;][]{2018PASA...35...10W} and the Stellar Abundances and Galactic Evolution Survey \citep[SAGES;][]{2023ApJS..268....9F}, yielding stellar-parameter estimates for about 50 million stars covering almost $3\pi$ steradians of sky \citep{2022ApJ...925..164H, 2023ApJ...957...65H}. To minimize the influence of molecular bands from carbon and nitrogen, the Pristine survey \citep{2017MNRAS.471.2587S} has proposed a narrower filter specifically centered on the wavelengths of the CaII H\&K doublet lines, hereafter referred to as the Ca HK-band\footnote{This technique was first explored by \citet{1991AJ....101.1902A,1995AJ....109.2828T,1998AJ....116.1922A,2000AJ....119.2882A}.}. Utilizing CFHT/MegaCam, the Pristine survey has observed over 6500 square degrees of the Northern sky with this specialized filter \citep{2023arXiv230801344M}.  Interestingly, the GALEX NUV-band is the most sensitive filter for  metallicity estimation excplored to date \citep{2024ApJS..271...26L}. Using GALEX and Gaia EDR3 data, \cite{2024ApJS..271...26L} have estimated metallicities with a precision of 0.11\,dex for about 4.5 million dwarfs and 0.17\,dex for approximately 0.5 million giants.

 On the other hand, metallicity can be precisely derived from broad-band filters with moderate sensitivity to metallicity (e.g., SDSS $u$) or very weak sensitivity but are observed with extremely high photometric precision (e.g., $G_{\rm BP}$).
For example, \cite{2015ApJ...803...13Y} determined metallicities for about 0.5 million stars with 1\% photometric precision 
from the SDSS Stripe 82 by \cite{2015ApJ...799..133Y}. \cite{2022ApJS..258...44X} acquired photometric metallicities with a typical precision of about 0.2\,dex for 27 million FGK stars from Gaia Early Data Release 3 \citep{2021A&A...649A...1G,2021A&A...650C...3G}. Using corrected Gaia DR3 XP spectra \citep{2024ApJS..271...13H}, B. Huang et al. (in preparation) synthesize Gaia 3-band photometry to estimate metallicities for approximately 100 million stars, with a precision  ($\sim$ 0.07\,dex) three times better compared to \cite{2022ApJS..258...44X}.

Although massive efforts have been carried out, the optimal design of filters for accurately deriving stellar parameters still remains to be fully explored. This important question is not easy to address, since the optimal design should consider not only the parameter sensitivity, but also the levels of uncertainty (including photon noise and calibration uncertainty) in the chosen filters, which limits the depth to which such surveys can be used. The latter is dependent on various factors, including the observing conditions and the survey strategies employed (e.g., the number of visits and the exposure times). 

As a first exploration, here we attempt to identify an optimal filter design for the derivation of stellar metallicity based on synthetic photometry from Gaia XP spectra \citep{2021A&A...652A..86C,2024ApJS..271...13H} convolved with a series of filter-transmission curves defined by different central wavelengths ($\lambda_{\rm c}$) and bandwidths ($\Delta\lambda$). This approach enables the identification of the highest-precision metallicity estimates in the $\lambda_{\rm c}$-$\Delta\lambda$ diagram, leading to optimal selections of $\lambda_{\rm c}$ and $\Delta\lambda$ for a given filter. The achieved precision is as good as $\sigma_{\rm [Fe/H]} \sim 0.03-0.04$\,dex, even better than that using the full information content of Gaia XP spectra \citep{2023ApJS..267....8A}.  Optimal filter designs for other stellar parameters (e.g., log\,$g$, [$\alpha$/Fe], [C/Fe]) can be determined in the same manner. Moreover, this analysis methodology can also be applied to any ground-based photometric surveys by simulating the parameter precision with synthetic photometry from theoretical spectra, with noise properly considered.

The organization of this letter is as follows.  Section \ref{sec:method} describes the data and methods employed. Results and a discussion are presented in Section \ref{sec:result}. Our conclusions are provided in Section \ref{sec:conclusion}.

\begin{figure*}[ht!]
   \centering
  \includegraphics[width=15.cm]{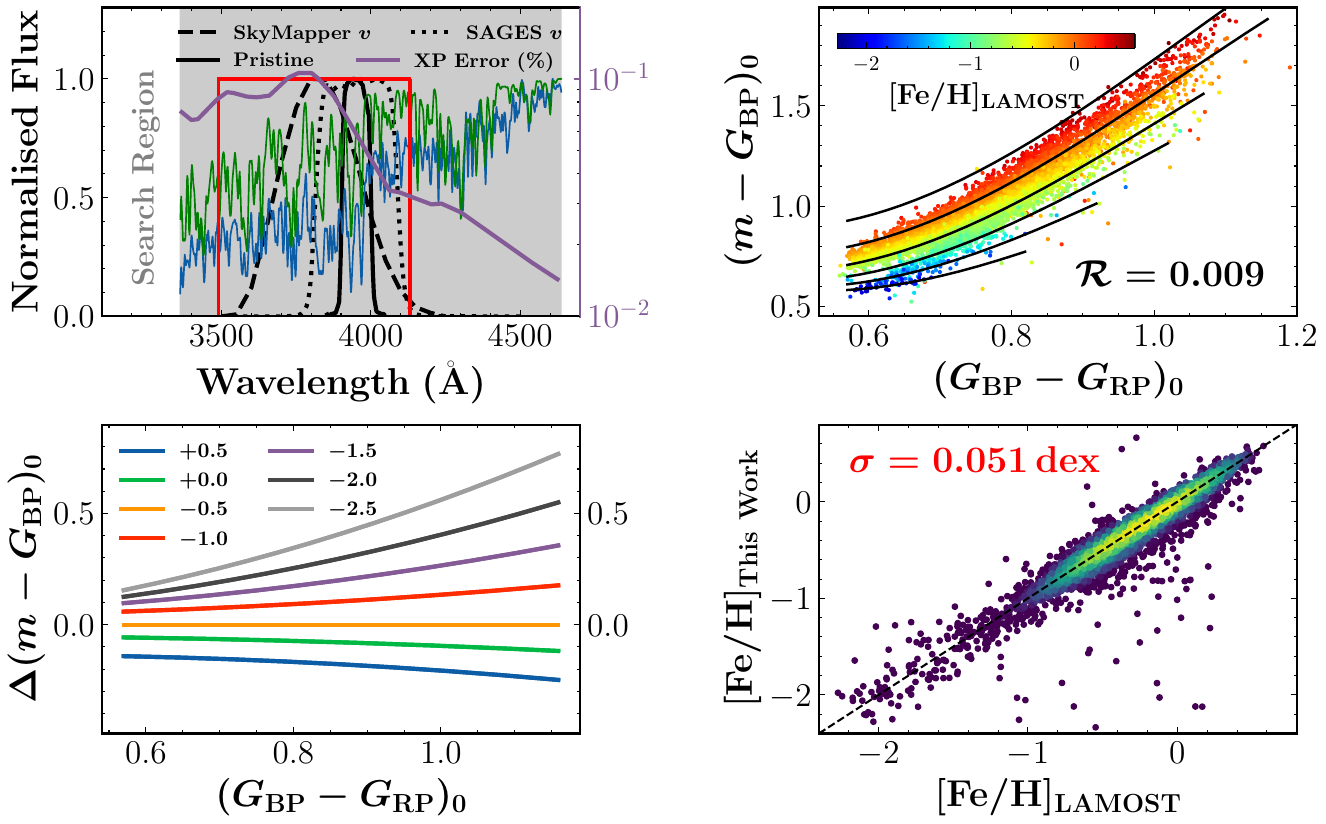}
   \caption{\small Upper-left panel: The gray-shadowed region (3360\,{\AA} to 4640\,{\AA}) represents the window over which we search to find the optimal design of a metallicity-sensitive filter.
   The red rectangle shows the region examined for the best design according to the algorithm described in Section\,2. Two synthetic spectra are plotted, one with $(T_{\rm eff}, {\rm log}\,g, {\rm [Fe/H]})=(5500\,{\rm K}, 4.5, 0.0)$ (blue line), and the other with $(5500\,{\rm K}, 4.5, -1.5)$ (green line), taken the from G{\"o}ttingen Spectral Library \citep{2013A&A...553A...6H}, in order to demonstrate the ``blanketing effect". The transmission curves of the SkyMapper $v$-band, SAGES $v$-band, and Pristine Ca\,HK-band are also shown. The dark-purple line represents the typical error of Gaia XP spectra taken from \citet{2024ApJS..271...13H} (see the green line in their Figure 8). Upper-right panel: Metallicity-dependent stellar loci of training-set stars in the plane of $(m-G_{\rm BP})_0$ and $(G_{\rm BP}-G_{\rm RP})_0$, color-coded by [Fe/H] as shown in the color bar. Here $m$ represents the synthetic magnitudes from the red-rectangle region in the upper-left panel. The black lines represent our best fits for [Fe/H], with values ranging from +0.5 (top) to $-2.0$ (bottom), in steps of 0.5\,dex, as described by Equation\,(1).
   The standard deviation of the fitting residuals, marked as $\mathcal{R}$ (in magnitudes), is labeled in the bottom-right corner.  Bottom-left panel: Variations of the stellar loci $\Delta (m-G_{\rm BP})_0$, as a function of $(G_{\rm BP}-G_{\rm RP})_0$, for different metallicities (as marked by different colors) relative to the locus at $\rm [Fe/H]=-0.5$. 
   Bottom-right panel: Comparison of the metallicity estimated from this work and LAMOST DR10. The standard deviation $\sigma_{\rm [Fe/H]}$ of the metallicity differences is marked in the top-left corner.}
  \label{Fig:f1}
\end{figure*}

\begin{figure*}[ht!]
   \centering
  \includegraphics[width=15.cm]{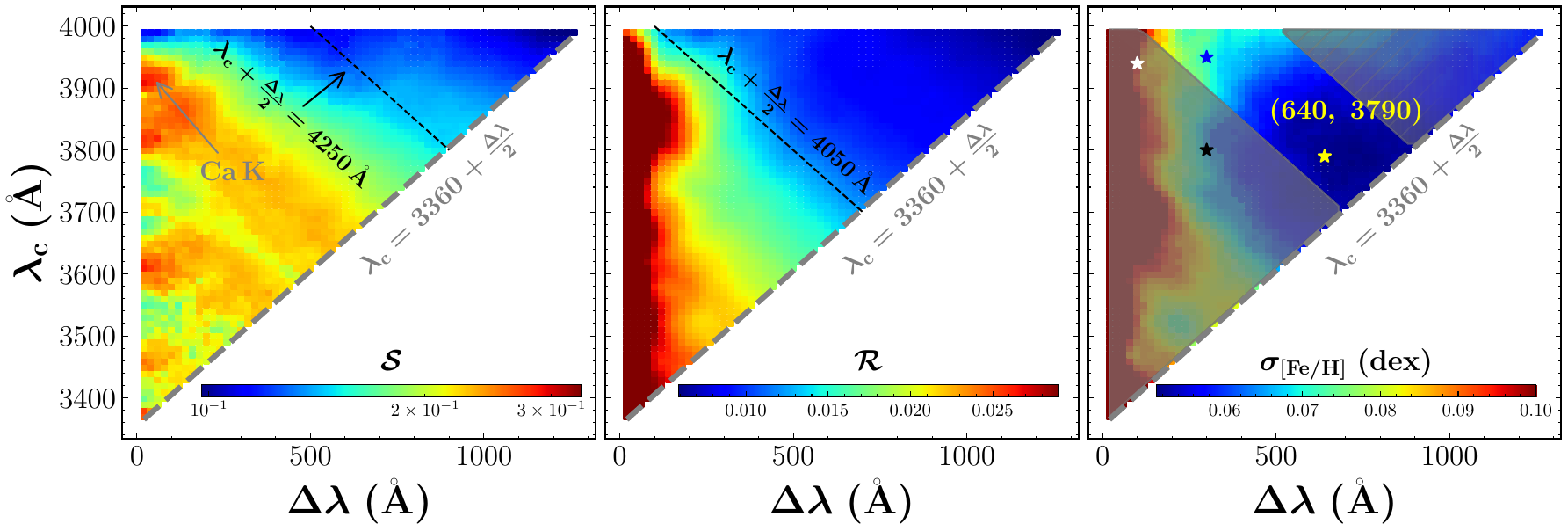}
   \caption{{\small Left panel: The metallicity sensitivity $\mathcal{S}$ in the $\lambda_{\rm c}$ and $\Delta \lambda$ plane. The sensitivity is significantly higher in the region with $\lambda_{\rm c} + \frac{\Delta \lambda}{2} \le 4250$\,{\AA}, indicated by the black-dashed line.
   The peak sensitivity centers around the wavelength of the Ca II K line (as indicated by the white arrow).
   A cut at $\lambda_{\rm c} \ge 3360+\frac{\Delta \lambda}{2}$ (gray-dashed line) is adopted, given the blue boundary of the Gaia XP spectra.
   Middle panel: Similar to the left panel, but for the scatter of the fitting residuals $\mathcal{R}$. The smallest scatter ($< 1$\%) tends to be in the region with $\lambda_{\rm c} + \frac{\Delta \lambda}{2} \ge 4050$\AA. Right panel: Similar to the left panel, but for the metallicity precision $\sigma_{\rm [Fe/H]}$. The shaded regions indicate the areas of poorer (largest) metallicity-precision estimate, while the non-shaded region indicates the better (lowest) metallicity-precision estimate. The yellow star marks the location of the optimal filter design with a central wavelength of 3790\,\AA\, and a width of 640\,\AA.  The black, blue, and white stars indicate the locations of the SkyMapper $v$-band, SAGES $v$-band, and Pristine Ca HK-band. respectively.
   }}
  \label{Fig:f0}
\end{figure*}

\section{Data and Methods} \label{sec:method}

\subsection{``Corrected'' Gaia XP Spectra} \label{sec:xp}

In addition to accurate astrometric information for over a billion stars, Gaia Data Release 3 \citep[DR3;][]{2023A&A...674A...1G} released ultra-low resolution ($\lambda/\Delta \lambda\sim$ 50) prism spectra covering extremely wide bands in the blue ($BP$-band) and red ($RP$-band), collectively referred to as XP spectra, for around 220 million stars, mostly with $G<17.65$. The XP spectral range spans from 336 to 1020\,nm. Recently, systematic errors in the XP spectra have been carefully re-calibrated by \cite{2024ApJS..271...13H}. By utilizing well-calibrated fluxed spectra from CALSPEC \citep{CALSPEC22}, NGSL \citep{NGSL},  and LAMOST \citep{2012RAA....12.1197C}, systematic offsets in the color, magnitude, reddening, and wavelength spaces have been thoroughly examined and corrected.  In this study, unless otherwise specified, the adopted Gaia XP spectra are the corrected version presented by \cite{2024ApJS..271...13H}.

\subsection{LAMOST Data Release 10} \label{sec:lm}

LAMOST is a 4\,m quasi-meridian reflecting Schmidt telescope equipped with 4000 fibers distributed over a field-of-view of 20 square degrees \citep{2012RAA....12.1197C}.  Starting from Phase-II, the LAMOST survey collected both low- ($R \sim 2000$) and medium-resolution optical spectra ($R \sim 7500$). In this study, we adopt the low-resolution data from LAMOST DR10, which has released over ten million low-resolution spectra, with reliable stellar-parameter estimates for about six million unique stars.  The stellar-atmospheric parameters used here are derived from the official pipeline -- the LAMOST Stellar Parameter Pipeline \citep[LASP;][]{2011RAA....11..924W, 2015RAA....15.1095L}. By using multiple observations, the internal precision is found to be 0.10, 0.06, 0.04 and 0.02\,dex for stars with spectral signal-to-noise ratio (SNR) of $10-20$, $20-40$, $40-80$, and greater than 80. As examined by \cite{2022ApJ...925..164H}, the LAMOST metallicity is robust over a wide range, with the metal-poor end extending to [Fe/H] $\sim -2.5$. For the purpose of deriving metallicities for the rare extremely metal-poor (EMP; [Fe/H]\,$\le -3.0$) stars, samples constructed with special pipelines appropriate for lower-metallicity stars are required \citep[see details in][]{2022ApJ...925..164H, 2023ApJ...957...65H}, beyond the scope of the present study.

\subsection{Training and Testing Sets} \label{sec:train}

The LAMOST DR10 catalog is cross-matched to the Gaia DR3 XP spectra with a distance radius of $1^{\prime\prime}$. In total, 5,571,175 dwarf stars are successfully matched.
To define the high-quality training sample, the following criteria are then applied:

\begin{enumerate}

\item[1)] Main-sequence stars with $M_{\rm G} = -(G_{\rm BP}-G_{\rm RP})^2 + 6.5\times(G_{\rm BP}-G_{\rm RP})-1.8$ and $4500 \le T_{\rm eff} \le 6500$\,K are selected;

\item[2)] To avoid contamination from binary stars, we require a Gaia renormalized unit error weight (RUWE) smaller than 1.2;

\item[3)] Spectral SNR within the LAMOST $g$-band $\geq 50$;

\item[4)] Distances from the Galactic plane $|Z| \ge 300$\,pc and $E (B -V) \le 0.01$ to minimize uncertainties due to reddening corrections.  The extinction values are taken from the dust map of \cite[][hereafter $E(B -V)_{\rm SFD}$]{1998ApJ...500..525S}, which is sufficient for stars in low-reddening regions. 

\end{enumerate}

After application of the above criteria, a total of 8107 stars remain. Around two-thirds (5107/8107) are selected as the training set. The remaining 3000 stars, along with 25,254 stars selected in the same way, but with SNR between 10 and 50, are adopted as the test set.

\subsection{Methods} \label{sec:ms}

The underlying physics to design a metallicity-sensitive filter is based on the so-called ``blanketing effect"  -- metals in stellar atmospheres mainly absorb radiation energy in the blue to near-ultraviolet region. This effect can be clearly seen over much of the entire blue region, shown in the upper-left panel of Figure\,\ref{Fig:f1}, until the wavelength approaches $\sim$ 4300\,{\AA}. We then explore the optimal design of a metallicity-sensitive filter in the following manner:

\begin{enumerate}

\item[1)] The transmission-curve filter is assumed to be have a top-hat shape, determined by two parameters: central wavelength $\lambda_{\rm c}$ and bandwidth $\Delta \lambda$. 
We tested different $\lambda_{\rm c}$ values, ranging between 3360\,$\rm {\AA}$ and 4000\,$\rm {\AA}$, in steps of 10\,$\rm {\AA}$. For each choice of $\lambda_{\rm c}$, $\Delta \lambda$ can vary from 20\,$\rm {\AA}$ to ($2\lambda_{\rm c}-3360)\,\rm {\AA}$, in steps of 20\,$\rm {\AA}$.
The reddening coefficient of each filter is derived by convolving its transmission curve with the Fitzpatrick extinction law assuming an $R_{V}=3.1$  \citep{1999PASP..111...63F}.

\item[2)] Convolve the Gaia XP spectra with the above-specified transmission curves to calculate the synthetic magnitudes, $\boldsymbol{m}$. Similarly, we predict synthetic magnitudes for the Gaia $G$, $G_{\rm BP}$, and $G_{\rm RP}$ bands. The reddening coefficient for $G_{\rm BP}-G_{\rm RP}$ is derived from \cite{2023ApJS..264...14Z}.

\item[3)] Similar to \citet{2015ApJ...803...13Y} and \citet{2022ApJ...925..164H}, 
metallicity-dependent stellar loci of the de-reddened colors $(m-G_{\rm BP})_0$ and $(G_{\rm BP}-G_{\rm RP})_0$ are defined in order to estimate stellar metallicity.  A third-order 2D polynomial (with 10 free parameters, including cross terms) is adopted to fit the color $(m - G_{\rm BP})_0$, as a function of $(G_{\rm BP}-G_{\rm RP})_0$ and [Fe/H], for the stars in the training set:
\begin{eqnarray}
 \begin{aligned}
  f(x, y)=~\sum_{j=0}^{3} \sum_{i=0}^{j}a_{i, j-i} \cdot x^{i} \cdot y^{j-i}~,
  \label{intrinsic_color_mod}
 \end{aligned}
\end{eqnarray}
where $x$ and $y$ refer to $(G_{\rm BP}-G_{\rm RP})_0$ and $\rm [Fe/H]$, respectively. It is worth noting that the optimal design may change slightly if adopting other broad-band filter systems (e.g., SDSS $g$- and $i$-bands).

An example of the fitting result is shown in the upper-right panel of Figure\,\ref{Fig:f1}; the standard deviation of the fitting residual is denoted by $\mathcal{R}$. The bottom-left panel of Figure \ref{Fig:f1} displays the variations in color $(m-G_{\rm BP})_0$ across different metallicities for different stars represented by different colors $(G_{\rm BP}-G_{\rm RP})_0$. Specifically, we define the metallicity sensitivity (denoted as $\mathcal{S}$) as the variation of $(m-G_{\rm BP})_0$ between $\rm [Fe/H] = +0.5$ and $\rm [Fe/H] = -0.5$ for Solar-type stars with $(G_{\rm BP}-G_{\rm RP})_0 = 0.80$.

\item[4)] Using the above empirical stellar loci represented by Equation\,\ref{intrinsic_color_mod}, we determine the stellar-metallicity estimate for each star in the test sample from its colors $(m-G_{\rm BP})_0$ and $(G_{\rm BP}-G_{\rm RP})_0$ using a numerical nonlinear equation root-finding algorithm (the Muller method; \citealt{Muller1956AMF}). A comparison between the photometric-metallicity estimates and the spectroscopic estimates obtained from LAMOST is shown in the bottom-right panel of Figure\,\ref{Fig:f1}. The standard deviation $\sigma_{\rm [Fe/H]}$ of the differences in metallicity, determined from Gaussian fitting, serves as the most important indicator (hereafter metallicity precision) for defining the optimal design of the metallicity-sensitive filter. 

\end{enumerate}

\begin{deluxetable}{ccccccccccc}[ht!]
\tablecaption{Metallicity sensitivities $\mathcal{S}$, Scatter of Fitting Residuals $\mathcal{R}$, and Metallicity Precision  $\sigma_{\rm [Fe/H]}$ for the Optimal Filter Design in this work and Three 
Custom Filters\\
 \label{tab:1}}
\tablehead{
\colhead{Filter} & \colhead{$\mathcal{R}$ (mag)} & \colhead{$\mathcal{S}$ (dex)} & \colhead{$\sigma_{\rm [Fe/H]}$ (dex)}}
\startdata
This work & $0.0094$ & $0.186$ & $0.051$ \\
SkyMapper $v$ & $0.0154$ & $0.247$ & $0.068$ \\
SAGES $v$ & $0.0140$ & $0.190$ & $0.075$ \\
Pristine Ca\,HK & $0.0208$ & $0.231$ & $0.108$ \\
Gaia broad bands & $0.0003$ & $0.004$ & $0.068$\\
\enddata
\end{deluxetable}

\begin{figure*}[ht!]
   \centering
  \includegraphics[width=16.cm]{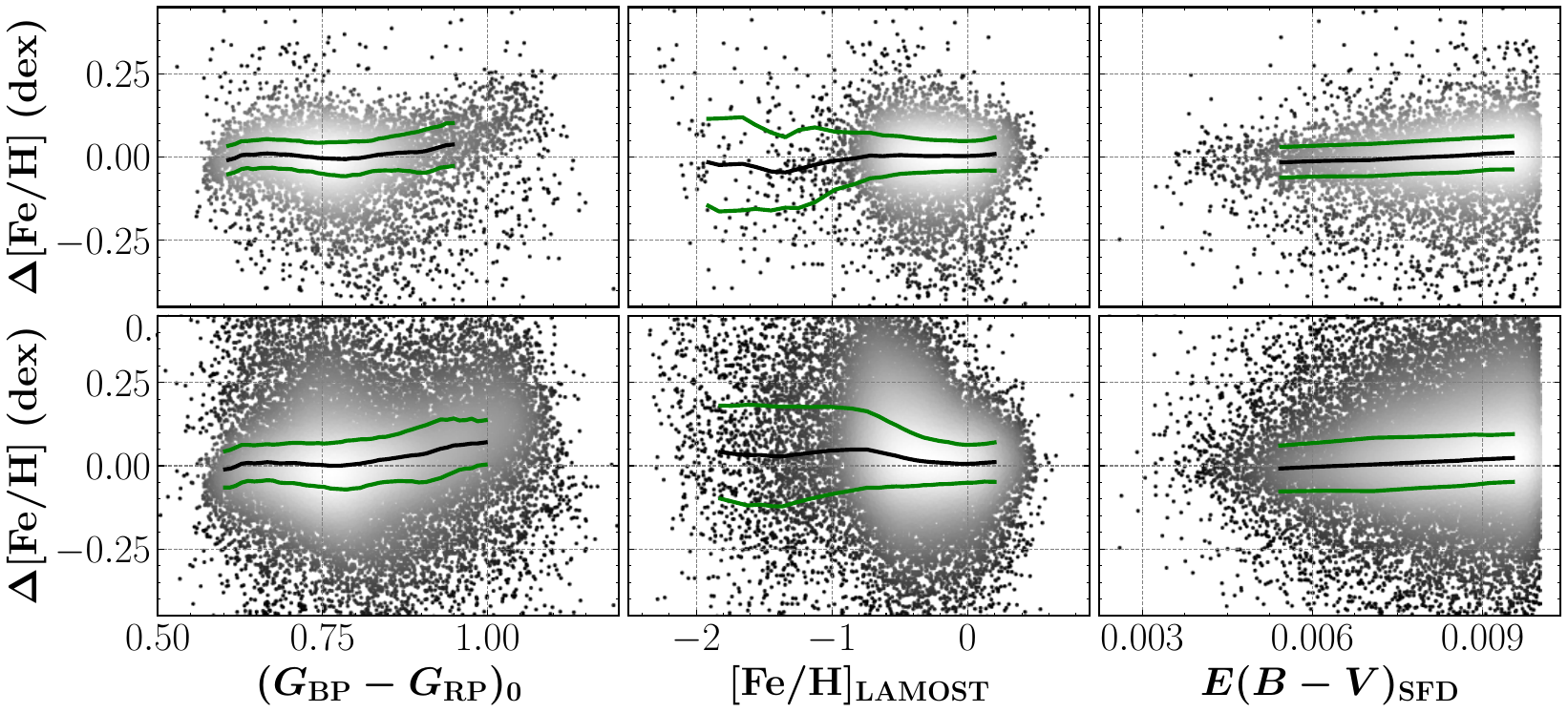}
   \caption{{\small Differences between the metallicity estimated from this work using the optimal metallicity-sensitive filter design (i.e., $\lambda_{\rm c}=3790\,\rm \AA,~\Delta \lambda=640\,\rm \AA$) and that estimated from LAMOST DR10, as a function of $(G_{\rm BP}-G_{\rm RP})_0$ (left), LAMOST $\rm [Fe/H]$ (middle), and $E(B-V)_{\rm SFD}$ (right) for both the training (upper) and testing (lower) samples. The black and green lines in each panel are the median and $1\sigma$ values from Gaussian fitting, respectively.}}
  \label{Fig:f2}
\end{figure*}

\begin{figure*}[ht!]
   \centering
  \includegraphics[width=9.cm]{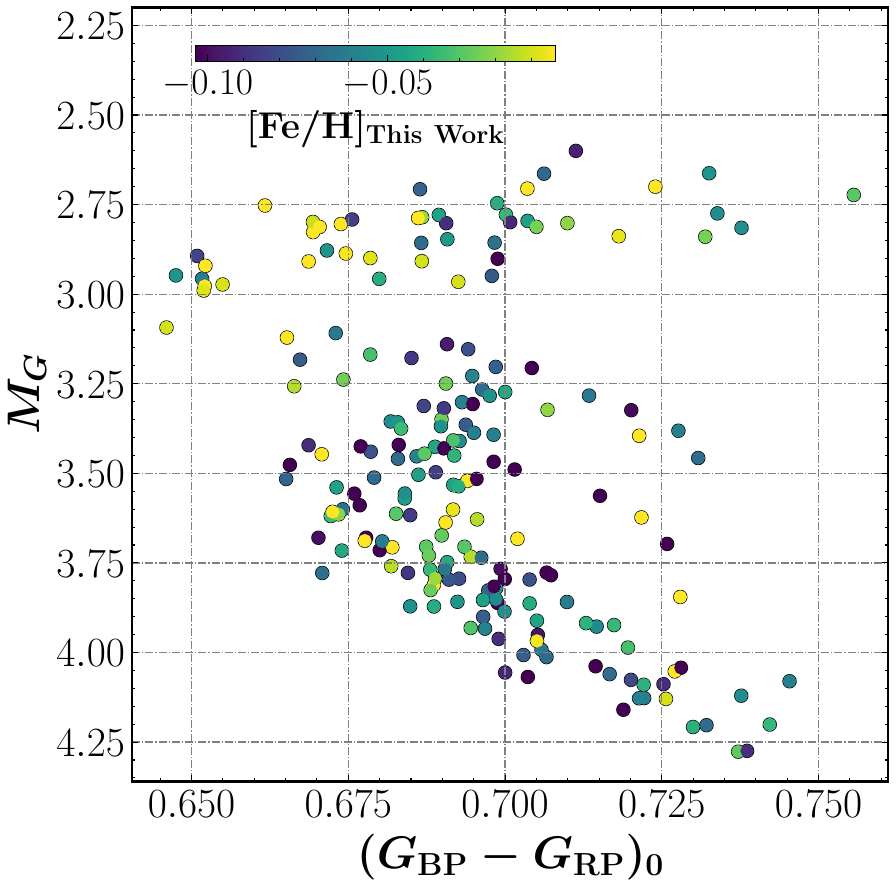}
  \includegraphics[width=8.7cm]{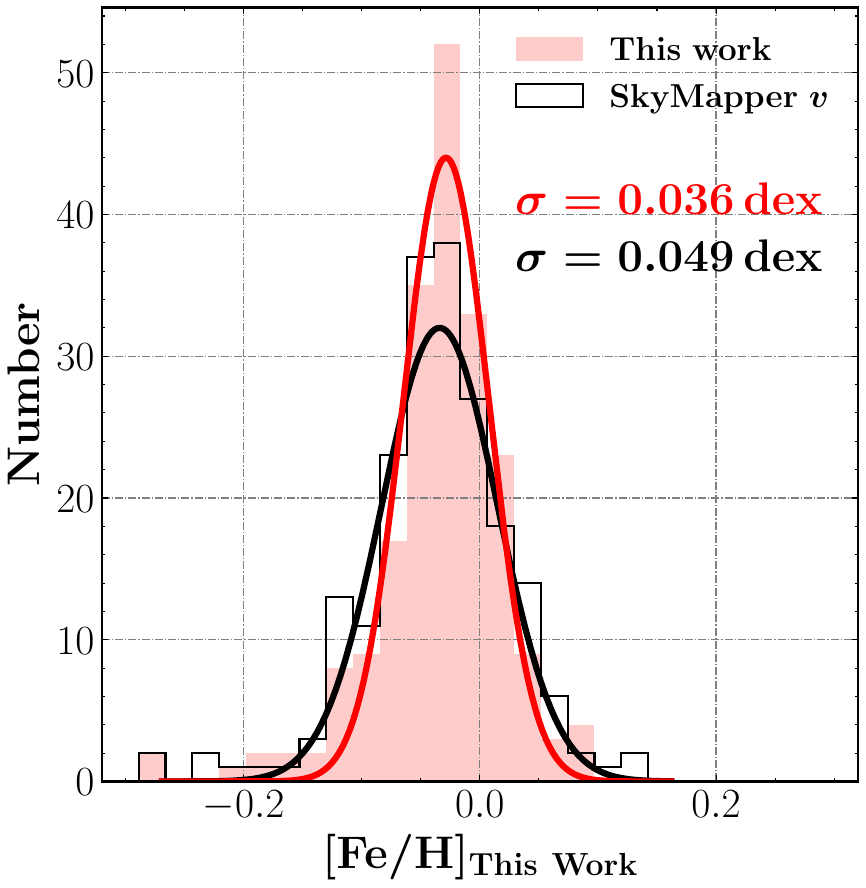} \\
   \caption{{\small External validation of the metallicity precision from member stars of the open cluster $\texttt{NGC 2682}$. Left panel: Color-absolute magnitude diagram of $\texttt{NGC 2682}$ member stars, color-coded by our photometric-metallicity estimate, as indicated in the top-left color bar. The giant branch stars are excluded by the cut: $M_{\rm G} \ge -(G_{\rm BP}-G_{\rm RP})_0^2 + 6.5\times(G_{\rm BP}-G_{\rm RP})_0-1.8$.  Right panel: Photometric metallicity estimated using the optimal filter design (blue) and the SkyMapper $v$-band (black line), respectively. The red and green curves are best-fit Gaussians for the metallicity distributions from our work and the SkyMapper $v$-band, respectively. The $1\sigma$ values from the fits are labeled in the upper-right corner.
   }}
  \label{Fig:f4}
\end{figure*}

\begin{figure*}[ht!]
   \centering
  \includegraphics[width=15.cm]{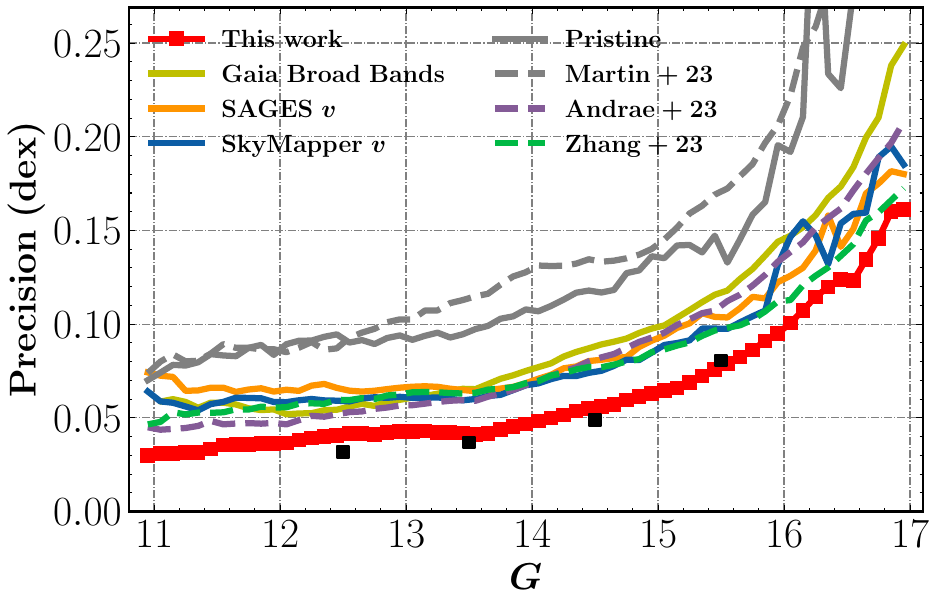}
   \caption{{\small Standard deviations (after subtracting the internal error of the LAMOST metallicities) of the difference between the photometric metallicity estimated from the optimal passband in this work (red line), SAGES $v$-band (saffron line), SkyMapper $v$-band (dark-blue line), Pristine Ca\,HK-band (light-blue line), and the LAMOST metallicity estimate, as a function of $G$-band magnitude. Similar results are shown comparing the literature 
   metallicities from  \cite{2023ApJS..267....8A} (purple-dashed curve), \cite{2023arXiv230801344M} (gray-dashed curve), \cite{2023MNRAS.524.1855Z} (green-dashed curve), and Gaia broad bands (B. Huang et al. in preparation) (yellow curve) to the LAMOST metallicity. The black boxes mark the theoretical precision predicted using the method in Appendix\,A. }}
  \label{Fig:f3}
\end{figure*}

\section{Results and Discussion} \label{sec:result}

Figure\,\ref{Fig:f0} illustrates the metallicity sensitivity $\mathcal{S}$, scatter of fitting residuals $\mathcal{R}$, and photometric precision $\sigma_{\rm [Fe/H]}$ for a series of filter designs characterized by the passband central wavelength $\lambda_{\rm c}$ and width $\lambda_{\rm c}$. As clearly shown in the plot, $\mathcal{S}$ is significantly enhanced for those filters with $\lambda_{\rm c} + \frac{\Delta \lambda}{2} \le 4250$\,{\AA}, except for several clumps at 3400\,{\AA}, 3530\,{\AA}, and 3750\,{\AA}, caused by the large errors in the Gaia XP spectra (indicated by the dark-purple line in the upper-left panel of Figure\,\ref{Fig:f1}). The peak sensitivity is located at 3910-3950\,{\AA}, corresponding to the wavelength of the CaII K line.  This location is also close to the central wavelength of most custom metallicity-sensitive filters (e.g., the SkyMapper $v$-band and Pristine Ca\,HK-band).  

On the other hand, the scatter of the fitting residuals $\mathcal{R}$ tends to be smaller than 1\% for those filters with $\lambda_{\rm c} + \frac{\Delta \lambda}{2} \ge 4050$\,{\AA}. This is because the uncertainty in the Gaia XP spectra significantly decreases from 10\% (at 3900\,{\AA}) to 2-3\% (at 4050\,{\AA}), as shown again by the dark-purple line in the upper-left panel of Figure\,\ref{Fig:f1}.

Finally, the precision of metallicity $\sigma_{\rm [Fe/H]}$ is determined by the combination of metallicity sensitivity $\mathcal{S}$ and the scatter of the fitting residuals $\mathcal{R}$.  As expected, the best precision is achieved by those filters with 4050\,{\AA}$ \le \lambda_{\rm c} + \frac{\Delta \lambda}{2} \le 4250$\,{\AA}.  Quantitatively, the optimal design is the filter with a central wavelength of $3790\,\rm \AA$ and a width of $640\,\rm \AA$, yielding an overall metallicity precision as good as 0.05\,dex, associated with the ratio of $\frac{\mathcal{R}}{\mathcal{S}}$. As a comparison, the well-known metallicity-sensitive filters generally have higher sensitivity $\mathcal{S}$ but larger scatter $\mathcal{R}$. The overall precision of those filters are all worse than the optimal one found in this work (see Table\,\ref{tab:1}).

We perform both internal and external checks to validate the precision of our derived photometric metallicity.  
First, the differences between the photometric-metallicity estimates and the LAMOST metallicity estimates, as a function of intrinsic color $(G_{\rm BP}-G_{\rm RP})_0$, LAMOST [Fe/H], and $E(B-V)_{\rm SFD}$, are shown in Figure\,\ref{Fig:f2}. No obvious systematic trends on $\rm [Fe/H]_{\rm LAMOST}$ and $E(B-V)_{\rm SFD}$ are found for both the training and testing samples. The dependence on $(G_{\rm BP}-G_{\rm RP})_0$ at $(G_{\rm BP}-G_{\rm RP})_0>0.9$ is also ignorable.

Secondly, member stars of open cluster $\texttt{NGC 2682}$ are selected to independently check the precision of our photometric-metallicity estmates for Solar-metallicity stars. We select $\texttt{NGC 2682}$ stars with membership probability $P_{\rm mem} \ge 0.9$ provided by \cite{2021ApJ...923..129J}. All of these member stars are required to be brighter than $G = 14$ and have Gaia XP spectra. By adopting a cluster distance of 1100\,pc \citep{2019ApJ...874...97S} and a reddening of $E(B-V) = 0.031$ from the SFD map, the absolute magnitudes $M_{G}$ and intrinsic color  $(G_{\rm BP}-G_{\rm RP})_0$ of these member stars are derived. A cut in the $M_{\rm G}$--$(G_{\rm BP}-G_{\rm RP})_0$ diagram is applied to exclude giant stars (see the left panel of Figure\,\ref{Fig:f4}). A total of 203 $\texttt{NGC 2682}$ members remain. As shown in the right panel of Figure\,\ref{Fig:f4}, a tiny dispersion of  $\sigma_{\rm [Fe/H]} = 0.036$\,dex is found for our photometric metallicities for those member stars. In contrast, the scatter of the photometric metallicities achieved by the well-known SkyMapper $v$-band (synthetically generated using the same method outlined in Section 2) is 0.049\,dex. The median photometric metallicity of [Fe/H] = $-0.038$ is very close to the value of [Fe/H] = $-0.031$ measured from high-resolution spectroscopy \citep{2019ApJ...874...97S}. 

Finally, using the testing sample of FGK dwarf stars ($4500 \le T_{\rm eff} \le 6500$\,K), Figure\,\ref{Fig:f3} shows the metallicity precision of our photometric-metallicity estimates, as a function of $G$-band magnitude. At the bright end, the uncertainty is only $\sigma_{\rm [Fe/H]} =
0.034$\,dex for $G < 11.5$, slightly increases to $\sigma_{\rm [Fe/H]} = 0.039$\,dex at $G = 12.0$, exhibits a plateau of $\sigma_{\rm [Fe/H]} = 0.044$\,dex for $12 < G < 14$, and quickly grows to $\sigma_{\rm [Fe/H]} > 0.1$\,dex towards the faint end at $G \sim 16.8$. Similarly, the metallicity precisions derived from custom metallicity-sensitive filters\footnote{Their synthetic photometry are also generated by convolving their transmission curves with Gaia XP spectra. As in Section\,2, we define the metallicity-dependent stellar loci for these filters from the training sample, and derive the photometric metallicities for stars in the testing sample.}, i.e.,  the SkyMapper $v$-band, Pristine 
Ca\,HK-band, SAGES $v$-band, and Gaia broad bands (B. Huang et al. in preparation), are shown for comparison. From inspection of the precisions listed in Table\,\ref{tab:1}, the results for the SkyMapper $v$-band, the SAGES $v$-band and Gaia broad bands are nearly identical, whereas those for the Pristine Ca\,HK-band is significantly worse; all are worse than the precision achieved by the optimal filter design in this study.
The four theoretical precision values at $G=12.5, 13.5, 14.5, 15.5$\,mag are also shown in Figure\,\ref{Fig:f3}, which are calculated using the method in the Appendix\,\ref{tpc}. Our estimates are consistent with the theoretical predictions.

Photometric-metallicity estimates derived from the synthetic Pristine Ca\,HK magnitudes, utilizing the Gaia XP spectra, are independently derived by \cite{2023arXiv230801344M}. Their results are cross-matched with the testing sample for calculating the uncertainty as a function of $G$-band magnitude, which is very close to that found by this work. In addition, the metallicities estimated from the entire Gaia XP spectra \citep{2023ApJS..267....8A, 2023MNRAS.524.1855Z}, utilizing machine-learning techniques, are compared to our testing sample. These results are also shown in Figure\,\ref{Fig:f3}; note that the internal uncertainty of the LAMOST metallicity estimates deduced from multiple observations has been subtracted from the above uncertainties. The Pristine and full XP precisions are all worse than our estimates across the entire magnitude range. This result confirms the core conclusion of this work -- the precision of the metallicity estimates (and presumably other stellar-atmospheric parameters) is the combined outcome of sensitivity and photometric uncertainty. 
Using this filter design, as well as Gaia broad-band photometry, we expect to achieve precise metallicity estimates for hundreds of millions of stars with Gaia XP spectra available. The precision is better than previous determinations by about a factor of two over all magnitude ranges.

\section{Conclusion} \label{sec:conclusion}

In this letter, we perform an intial exploration of the optimal filter design for deriving stellar metallicity, considering not only the metallicity sensitivity but also the uncertainty, including both calibration error and photon noise. Specifically, a series of filters with top-hat transmission curves (determined by central wavelength $\lambda_c$ and passband width $\Delta \lambda$) are convolved with Gaia XP spectra. Using stars with available LAMOST spectroscopic metallicities as training and testing samples, the optimal filter design (providing the best metallicity precision) is found at ($\lambda_c$, $\Delta \lambda$) = 
(3790, 640)\,{\AA}, which is a combined result of metallicity sensitivity and uncertainty.

An internal check with metallicity estimates of LAMOST FGK dwarf stars ($4500 \le T_{\rm eff} \le 6500$\,K) shows that the precision is as good as $\sigma_{\rm [Fe/H]} = 0.034$\,dex at the bright end ($G \leq 11.5$), and is better than $\sigma_{\rm [Fe/H]} = 0.15$\,dex at the faint end, with $G \sim 17$.  These precisions are superior to those achieved by well-known custom filters such as the SkyMapper and SAGES $v$-band, the Pristine Ca\,HK-band, and even by the use of the entire XP spectra. An external test with over two hundred member stars of the roughly Solar-abundance open cluster {\tt NGC 2682} reveals a tiny scatter of $\sigma_{\rm [Fe/H]} = 0.036$\,dex in metallicity, in excellent agreement with our internal check.

The method presented in this paper can be used to yield the most precise measurements of stellar elemental abundances  (e.g., metallicity, carbonicity, and the alpha elements) from XP spectra, and offer critical insights into filter design for large-scale survey projects. For instance, one can estimate the SNR at each wavelength by considering various factors, such as the telescope’s aperture, exposure time, atmospheric extinction at the observatory, seeing, the quantum efficiency of the detector, and the transmission function of the telescope's optical system. Utilizing the model spectra, this approach enables the prediction of the optimal design of filters for estimation of various stellar parameters. The result of this work also highlights the need to achieve mmag level calibration precision in near-UV filters.


\begin{acknowledgments}

This work is supported by the National Key Basic R\&D Program of China via 2023YFA1608300; the China Postdoctoral Science Foundation of 2023M743447; the National Natural Science Foundation of China through the project NSFC 12222301, 12173007 and 11603002, the National Key Basic R\&D Program of China via 2019YFA0405503 and Beijing Normal University grant No. 310232102. 
We acknowledge the science research grants from the China Manned Space Project with NO. CMS-CSST-2021-A08 and CMS-CSST-2021-A09. T.C.B. acknowledges acknowledge partial support for this work from grant PHY 14-30152; Physics Frontier Center/JINA Center for the Evolution of the Elements (JINA-CEE), and OISE-1927130: The International Research Network for Nuclear Astrophysics (IReNA), awarded by the US National Science Foundation.  

This work has made use of data from the European Space Agency (ESA) mission
Gaia (\url{https://www.cosmos.esa.int/gaia}), processed by the Gaia
Data Processing and Analysis Consortium (DPAC,
\url{https://www.cosmos.esa.int/web/gaia/dpac/consortium}). Funding for the DPAC
has been provided by national institutions, in particular the institutions
participating in the Gaia Multilateral Agreement.
Guoshoujing Telescope (the Large Sky Area Multi-Object Fiber Spectroscopic Telescope LAMOST) is a National Major Scientific Project built by the Chinese Academy of Sciences. Funding for the project has been provided by the National Development and Reform Commission. LAMOST is operated and managed by the National Astronomical Observatories, Chinese Academy of Sciences.

\end{acknowledgments}

\clearpage
\appendix
\setcounter{table}{0}   
\setcounter{figure}{0}
\renewcommand{\thetable}{A\arabic{table}}
\renewcommand{\thefigure}{A\arabic{figure}}
\section{theoretical precision calculation}
\label{tpc}

Assuming the covariance of spectral errors is negligible, the theoretical precision of predicted metallicity using the optimal filter can be computed by considering the photometric uncertainty and sensitivity to metallicity. This can be calculated using the equation:
\begin{eqnarray}
 \begin{aligned}
    \sigma_{\rm [Fe/H]} =& \frac{\sqrt{\sum_{\lambda} (\frac{\mathcal{E}(\lambda) \cdot F(\lambda) \cdot T(\lambda)}{h \nu})^2}}{\sum_{\lambda} \frac{\mathcal{G}(\lambda) \cdot F(\lambda) \cdot T(\lambda)}{h \nu}}\text{,}
 \end{aligned}
\end{eqnarray}   
where ${\mathcal E}(\lambda)$, $F(\lambda)$, $T(\lambda)$, ${\mathcal G}(\lambda)$, and $h\nu$ denote the relative error, flux, transmission function of the optimal filter, gradient spectra at wavelength $\lambda$, and energy of a photon, respectively. The relative errors of Gaia XP spectra are ratios of the flux error to the flux, both of which are derived 
by $\texttt{GaiaXPy}$ \citep{2022zndo...6674521R}. Based on the same testing sample, we further compute the ratio of the mean normalized spectra for $\rm [Fe/H]_{LAMOST}$ ranging from -0.8\,dex to +0.0\,dex in increments of 0.1\,dex, to the spectra corresponding to $\rm [Fe/H]_{LAMOST}=-0.8$\,dex. Subsequently, the gradient spectrum is estimated as the mean rate of change of this ratio with respect to $\rm [Fe/H]_{LAMOST}$. Note that the yielded gradient spectrum is consistent with the result of the 54 mini-JPAS narrow band filters \citep{2023MNRAS.518.2018Y}.

\end{document}